\documentclass[namedreferences]{solarphysics}
\usepackage[hyperref,optionalrh,solaromanenum]{spr-sola-addons} 
\usepackage{epsfig}                     
\usepackage{graphicx}                    
\usepackage{color}                       
\usepackage{breakurl}                         
\usepackage{amssymb}

\usepackage[T1]{fontenc}

\newcommand{\etal}{et al.}

\newcommand{\pder}[2]{ \frac{\partial #1}{\partial #2} }
\newcommand{\pderN}[3]{ \frac{\partial^{#3} #1}{\partial #2^{#3}} }

\chardef\us=`\_

\begin{document}

\begin{article}

\begin{opening}
	
\title{Dispersion of Slow Magnetoacoustic Waves in the Active Region Fan Loops Introduced by Thermal Misbalance }

%
\author[addressref={aff1,aff2},corref,email={mr\_beloff@mail.ru}]{\inits{S.A.}\fnm{S.}~\lnm{Belov}}
\author[addressref={aff2},corref,email={}]{\inits{N.E.}\fnm{N.}~\lnm{Molevich}}
\author[addressref={aff1,aff2},corref,email={}]{\inits{D.I.}\fnm{D.}~\lnm{Zavershinskii}}

%

\runningauthor{S. Belov \etal}
\runningtitle{Dispersion of Slow Magnetoacoustic Waves Introduced by Thermal Misbalance }

\address[id=aff1]{Department of Physics, Samara National Research University, Moscovskoe sh. 34, Samara, 443086, Russia}
\address[id=aff2]{Department of Theoretical Physics, Lebedev Physical Institute, Novo-Sadovaya st. 221, Samara, 443011, Russia}

\begin{abstract}

Slow magnetoacoustic waves observed in the solar corona {are used as seismological probes} of  plasma parameters. It has been shown that dispersion properties of such waves can vary significantly under the influence of the wave-induced thermal misbalance. In the current research, we study the effect of misbalance on waves inside the magnetic-flux tube under the second-order thin-flux-tube approximation. Using the parameters of active region fan coronal loops, we calculated wave properties such as the phase speed and decrement. It is shown that neglecting thermal misbalance may be the reason for the substantial divergence between seismological and spectrometric estimations of plasma parameters. We also show that the frequency dependence of phase speed is affected by two features, namely the geometric dispersion and the dispersion caused by the thermal misbalance. In contrast to the phase speed, the wave decrement primarily is affected  by the thermal misbalance only. The dependencies of the phase speed and decrement of the slow wave on magnetic field and tube cross-section are also analyzed.

\end{abstract}


\end{opening}

\section{Introduction}
Various magnetohydrodynamic waves are omnipresent in the solar corona (see, e.g., \cite{Nakariakov2020} for a recent review). In particular, many observed phenomena are associated with propagating or standing slow waves. For example, strongly damped Doppler-shift oscillations observed with the \textit{Solar Ultraviolet Measurements of Emitted Radiation} (SUMER) spectrometer onboard the \textit{Solar and Heliospheric Observatory} (SOHO) mission  and  commonly known as SUMER oscillations are attributed  to standing  slow  waves in hot coronal loops (\citealp{Wang2011}). Sloshing oscillations, which are observed as localized enhancements of the EUV emission intensity in hot coronal loops, bouncing back and forth between the footpoints (\citealp{Reale2016, Pant2017}), are associated with the evolution of slow modes in the closed magnetic configuration. Propagating slow waves are usually detected as quasi-periodic EUV and soft X-ray intensity perturbations. These perturbations propagate upward along field-aligned plasma non-uniformities and are detected in many coronal regions, for example, in active-region fan-like loops (\citealp{Yuan2012}), and in plume and inter-plume regions in polar coronal holes ({\citealp{Gupta2010}}). 

Propagating slow waves can be used to probe the direction (\citealp{Marsh2009}) and value (\citealp{Jess2016}) of the coronal magnetic field, as well as the effective polytropic index of a plasma (\citealp{Van_Doorsselaere2011}). The main feature of observed propagating slow waves is their quick decay with the damping lengths  $\approx 10\,\mathrm{Mm}$. Using spectroscopic and stereoscopic-imaging observations of slow-magnetoacoustic-wave propagation within a coronal loop, \cite{Marsh2011} investigated the damping of the slow magnetoacoustic mode in three dimensions. According to their analysis, the damping length is found to be of $20\,\mathrm{Mm}$. {The authors' subsequent forward modeling suggest} that the thermal conduction is insufficient to explain this value, given the observed
parameters of the coronal-loop temperature, density, and wave  period. In addition, the observed damping lengths are shorter for shorter wave periods (\citealp{Prasad2014}) and show no obvious decrease with temperature (\citealp{Prasad2019}).  Such features indicate that an alternative or additional dispersive/dissipative mechanism should be taken into consideration.

It was shown by \cite{Zavershinskii2019} and \cite{2019A&A...628A.133K} that the effect called thermal misbalance can significantly affect dispersion properties of slow waves in the solar corona. Thermal misbalance is a manifestation of the feedback between compression waves and heating/cooling processes operating in the plasma. This can lead to  the dependence of the phase speed and the rate of growth/decay on the period, cause an increase or additional damping of waves and lead to the formation of quasi-periodic patterns (see, e.g., \citealp{Zavershinskii2019}). Moreover, the effect of thermal misbalance can be associated with the observed temperature dependence of the polytropic index (see, e.g., \citealp{2018ApJ...868..149K}). The misbalance is also shown to be responsible for the phase shift between perturbations of various plasma parameters (density, temperature, etc.) and for the distribution of energy in and between eigenmodes, e.g. slow and entropy waves (see \citealp{2021arXiv210412652Z}). Particularly, the role of thermal misbalance in the estimation of the phase shifts is found to be significant for the high-density and low-temperature
loops. \cite{2021arXiv210407604P} show that the variation of heating mechanism may lead to around a five-fold increase in the phase difference. The previously mentioned behaviour of damping length can be explained by the influence of thermal misbalance on the slow-wave dynamics as well. \cite{Duckenfield2020effect} found that the damping times of slow waves due to thermal misbalance are of the order of $10$\,--\,$100$ minutes, which coincides with the wave periods and damping times observed. 

In this article, we will focus on slow waves propagating in active region fan loops. It is known from pioneering works by   \cite{1975IGAFS..37....3Z,1982SvAL....8..132Z} and \cite{1983SoPh...88..179E} that slow waves are a subject of dispersion due to the finite size of the wave-guiding structure. As a result of such a geometric dispersion, the phase speed of slow waves lies between the tube- and sound speeds inside the wave guide.  We aim to consider the combined influence of the finite cross section and thermal misbalance on slow MA waves. It will be shown that to describe the  propagation of slow waves in magnetic flux tubes of thermally active plasma, it is necessary to use a tube speed obtained with both geometric dispersion and dispersion due to thermal misbalance taken into account.

The organization of our article is as follows: in Section \ref{s:DispRealation}, one may find the evolutionary equation for compression modes propagating in a magnetic-flux tube composed of thermally active plasma. Using the equation obtained, we derive the relations describing dispersion properties of slow MA waves. In order to apply the  obtained  theoretical results to the coronal conditions, we specify the considered heat-loss model in Section \ref{ss:HLmodel}. Further, in Section \ref{ss:PhaseSpeed}, we turn to the dependence of the phase velocity on the period of the slow wave and compare our results with the results for an ideal plasma without thermal misbalance. We show that neglecting the impact from the heating/cooling process may be the source of significant errors in the seismological estimation of plasma parameters (see Section \ref{ss:Errors}). However, these deviations may be used for seismological determination of unknown coronal-heating mechanism as a solution of the reverse problem. In addition, in Section \ref{s:RoleofAB}, we analyze the role of the  cross-section scale and magnetic-field strength on dispersion properties of slow waves. The summary and discussion of the results presented can be found in the final section of this article.

\section{Dispersion Relation}\label{s:DispRealation} 
The general approach of \cite{1975IGAFS..37....3Z,1982SvAL....8..132Z} and \cite{1983SoPh...88..179E} gives extensive information about properties of MHD waves propagating inside a magnetic-flux tube. However, it is not always convenient for analysis. Therefore, we will use the second-order thin-flux-tube approximation (\citealp{Zhugzhda96}), which allows us to reduce the two-dimensional
consideration to a one-dimensional one. This assumption implies that $R\ll\lambda$, where $R$ is the tube radius and $\lambda$ is the characteristic length for density, velocity and other physical parameters variations along the tube. Further, we will consider MHD waves propagating in the magnetic-flux tubes composed of thermally active plasma. In order to take into account the thermal misbalance effect, the system of equations from \cite{Zhugzhda96} have been slightly modified by adding the heating and cooling rates in the right-hand side of the energy transport Equation \ref{energy}. Thus, the basic set of  equations is written as follows:

\begin{equation} \label{system_eq_first}
\pder{\rho}{t}+2\rho V +\pder{\rho u}{z}=0,
\label{Cont}
\end{equation}
\begin{eqnarray}
p+\frac{b_z^2}{8\pi}-\frac{A}{2\pi}\left[\rho\left(\pder{V}{t}+u\pder{V}{z}+V^2-\Omega^2\right)\right.+\nonumber\\
+\left.\frac{1}{4\pi}\left(J^2-\frac{1}{4}\left(\pder{b_z}{z}\right)^2+\frac{1}{2}b_z\pderN{b_z}{z}{2}\right)\right]=p_\mathrm{ext},
\label{motionV}
\end{eqnarray}
\begin{equation}
\rho\pder{u}{t}+\rho u\pder{u}{z}=-\pder{p}{z},
\label{motionU}
\end{equation}
\begin{equation}
\rho\pder{\Omega}{t}+\rho u\pder{\Omega}{z}+2\rho\Omega V=\frac{1}{4\pi}\left(b_z\pder{J}{z}-J\pder{b_z}{z}\right),
\label{motionOmega}
\end{equation}
\begin{equation}
\pder{J}{t}+\pder{uJ}{z}+2VJ-b_z\pder{\Omega}{z}=0\,,
\label{inductionJ}
\end{equation}
\begin{equation}
\pder{b_z}{t}+u\pder{b_z}{z}+2Vb_z=0\,,
\label{inductionBz}
\end{equation}
\begin{equation}
\pder{A}{t}+u\pder{A}{z}-2VA=0\,,
\label{crossSection}
\end{equation}
\begin{equation}
C_{V}\rho\left(\pder{T}{t}+u\pder{T}{z}\right)-\frac{\mathrm{k_B} T}{m}\left(\pder{\rho}{t}+u\pder{\rho}{z}\right)=-\rho Q\left(\rho,T\right),
\label{energy}
\end{equation}
\begin{equation} \label{system_eq_last}
p=\frac{\mathrm{k_B}}{m}\rho T.
\label{State}
\end{equation}

In Equations \ref{Cont}\,--\,\ref{State}, $\rho$, $T$, and $p$ are the density, temperature, and pressure of
the plasma, respectively; $u$ and $b_z$ are the plasma velocity and magnetic field along the flux tube, $V$ is the radial derivative of the radial velocity; $J$ and $\Omega$ are the current density and vorticity, respectively; $A=\pi R^2$ is the flux-tube cross section, $R$ is the tube radius; $p_\mathrm{ext}$ is the total (gas-dynamic and magnetic) external pressure ;  $\mathrm{k_B}$ is the Boltzmann constant; $C_{V}$ is the specific heat capacity at constant volume; $m$ is the mean mass per one particle. We use  $Q\!\left(\rho, T\right)=L\!\left(\rho, T\right)-H\!\left(\rho, T\right)$ for the heat-loss function (\citealp{Parker1953, Field1965}), where  $H\!\left(\rho, T\right)$ and $L\!\left(\rho, T\right)$ are  heating and radiation cooling rates, respectively. The heat-loss  function equals zero under steady-state condition: $Q\!\left(\rho_0, T_0\right)=L\!\left(\rho_0,T_0\right)-H\!\left(\rho_0, T_0\right)=L_0-H_0=0$. 

As the cooling and heating rates depend on density and temperature, the wave-induced perturbations of these rates cause the misbalance between non-adiabatic processes, which, in turn, affects waves. The impact of this feedback varies depending on the wave frequency $\omega$. Using characteristic  timescales $\tau_V=C_{V}/Q_{0T}$ and $\tau_P=C_{P} T_0/\left( Q_{0T}T_0-Q_{0\rho}\rho_0 \right)$, one may introduce ranges of weak ($\omega \left|\tau_{V,P}\right|\gg1$) and strong ($\omega \left|\tau_{V,P}\right|\ll1$) impacts of the thermal misbalance. Here, $C_{P}=C_{V}+\mathrm{k_B}/m$ is the specific heat capacity at constant pressure,  $Q_{0T}=\left.\partial Q/\partial T\right|_{\rho_0, T_0}$, $Q_{0\rho}=\left.\partial Q/\partial\rho\right|_{\rho_0, T_0}$.  It has been previously shown that  the phase speed $\left[c_\mathrm{ph}\right]$ of slow MA modes becomes frequency-dependent in the uniform plasma with thermal misbalance and varies between  $c_\mathrm{S}$ and $c_{\mathrm{S}Q}$ (see \citealp{Molevich88,Zavershinskii2019}). The harmonics weakly affected by the thermal misbalance ($\omega \left|\tau_{V,P}\right|\gg1$) propagate with  $c_\mathrm{S}=\sqrt{C_{P}\mathrm{k_B}\!T_0/C_{V}m}$, which is the standard value for the plasma without the thermal misbalance.  In the opposite case  ($\omega \left|\tau_{V,P}\right|\ll1$),  the heating and cooling processes fully determine the speed of sound {$c_{\mathrm{S}Q} =\sqrt{\tau_V C_{P}\mathrm{k_B} \!T_0/\tau_P C_{V}m}$}. The maximum of dispersion effect (where $\mathrm{d} c_\mathrm{ph}/\mathrm{d} \omega$ reaches maximum) is reached near the period
\begin{equation} \label{Per_mis}
	P_\mathrm{m} = 2\pi (\tau_{P}\tau_{V})^\frac{1}{2}.
\end{equation}

Let us analyze waves in a untwisted ($J_0=0$) and non-rotating ($\Omega_0=0$) flux tube. The linearization procedure applied to  Equations \ref{system_eq_first}\,--\,\ref{system_eq_last} allows us to obtain evolutionary equations for plasma eigenmodes. In the considered case, the Alfvén waves are not coupled with compressional modes and, therefore, can be excluded from our analysis. Focusing on compressional waves only and excluding all variables except the density perturbation $\rho_1$, gives us the following equation: 
\begin{equation} \label{waveEq}
	\pder{}{t}\hat{D} \rho_1+\frac{1}{\tau_V}\hat{D}_Q\rho_1=0,
\end{equation}
where
\begin{eqnarray}
\hat{D}=\left(\left(c_\mathrm{A}^2+c_\mathrm{S}^2\right)\left(\pderN{}{t}{2}-c_\mathrm{T}^2\pderN{}{z}{2}\right)+\frac{A_0}{4\pi}\left(\pderN{}{t}{2}-c_\mathrm{S}^2\pderN{}{z}{2}\right)\left(\pderN{}{t}{2}-c_\mathrm{A}^2\pderN{}{z}{2}\right)\right),\nonumber\\
\hat{D}_Q=\left(\left(c_\mathrm{A}^2+c_{\mathrm{S}Q}^2\right)\left(\pderN{}{t}{2}-c_{\mathrm{T}Q}^2\pderN{}{z}{2}\right)+\frac{A_0}{4\pi}\left(\pderN{}{t}{2}-c_{\mathrm{S}Q}^2\pderN{}{z}{2}\right)\left(\pderN{}{t}{2}-c_\mathrm{A}^2\pderN{}{z}{2}\right)\right).\nonumber
\end{eqnarray}

The evolutionary Equation \ref{waveEq} is the PDE of the fifth order in time $t$ and fourth order in coordinate $z$. We use notations $A_0$ for the unperturbed cross section and $c_\mathrm{A}^2=B_0^2/4\pi\rho_0$ for the square of Alfvén speed, where $B_0$ is the unperturbed magnetic field along the tube. It follows from Equation~\ref{waveEq} that for a loop with some specified temperature, the wave dynamics will be governed by loop geometry (through $A_0$), magnetic-field strength (through $c_\mathrm{A}^2$) and heating/cooling processes (through $ \tau_P, \tau_V$).   It can be shown that applying the infinite field approximation ($\beta \rightarrow 0$ and $A_0 k^2 \rightarrow 0$), Equation~\ref{waveEq} is reduced to the equation for compressional perturbations previously analyzed by \cite{Zavershinskii2019}. In the case of the absence of thermal misbalance $ \tau_{P,V} \rightarrow \infty $,  Equation~\ref{waveEq} transforms to the evolutionary equation for an ideal plasma \citep{Zhugzhda96}. 

It can be seen that the multiplier  $A_0$ is present in both of the operators $\hat{D}$ and $\hat{D}_Q$. This fact implies that the geometric dispersion will act not only at temporal scales weakly affected by the thermal misbalance ($\omega \left|\tau_{V,P}\right|\gg1$, see  $\hat{D}$), but also at frequencies that are strongly affected by the thermal misbalance  ($\omega \left|\tau_{V,P}\right|\ll1$, see  $\hat{D}_Q$). As a result,  to describe  the properties of waves, one should introduce not only the well-known  tube speed  $c_\mathrm{T} = \sqrt{c_\mathrm{A}^2c_\mathrm{S}^2/\left(c_\mathrm{A}^2+c_\mathrm{S}^2\right)}$, which is the consequence of single geometric dispersion, but also the modified tube-speed value following from the combination of geometric and thermal-misbalance dispersion effects:
\begin{equation}
	\label{cT0}
	c_{\mathrm{T}Q}= \sqrt{\frac{c_\mathrm{A}^2c_{\mathrm{S}Q}^2}{c_\mathrm{A}^2+c_{\mathrm{S}Q}^2}}.
\end{equation}
 
As we are interested in the dispersion properties of waves, we search  for the solution for Equation~\ref{waveEq} of the form $\mathrm{e}^{-\mathrm{i}\omega t + \mathrm{i} kz}$. Collecting  together terms involving the same powers of $k$, we can write the dispersion relation  as:
\begin{equation}
ak^4+bk^2-c=0, \label{dispRelation}
\end{equation}
where
\begin{eqnarray}
a=\frac{A_0}{4\pi}\left(c_{\mathrm{S}Q}^2-\mathrm{i}\omega\tau_V c_\mathrm{S}^2\right)c_\mathrm{A}^2,\nonumber \qquad\qquad\qquad\qquad\qquad\\
b=\left(\left(c_\mathrm{A}^2+c_{\mathrm{S}Q}^2\right)\left(c_{\mathrm{T}Q}^2-\frac{A_0}{4\pi}\omega^2\right)-\mathrm{i}\omega\tau_V\left(c_\mathrm{A}^2+c_\mathrm{S}^2\right)\left(c_\mathrm{T}^2-\frac{A_0}{4\pi}\omega^2\right)\right),\nonumber\\
c=\left(\left(\left(c_\mathrm{A}^2+c_{\mathrm{S}Q}^2\right)-\frac{A_0}{4\pi}\omega^2\right)-\mathrm{i}\omega\tau_V\left(\left(c_\mathrm{A}^2+c_\mathrm{S}^2-\frac{A_0}{4\pi}\omega^2\right)\right)\right)\omega^2.\nonumber
\end{eqnarray}
The  dispersion relation (Equation \ref{dispRelation}) describes five compressional modes including two fast and two slow MA waves, and also one entropy mode. Considering the first order thin flux tube approximation $A_0 k^2 \rightarrow 0$, reduced dispersion relation~\ref{dispRelation} coincides with those presented in \cite{Duckenfield2020effect} (assuming the weak impact of thermal conduction and no dependence of the heating rate on magnetic field strength). The  entropy mode is the non-propagating mode with the real part of the frequency identically equal zero ($\mathrm{Re}~\omega = 0$) independently of the heat-loss function. It should be mentioned that in some specific regimes of thermal misbalance (specific forms of the heat-loss function), the slow-mode harmonics can demonstrate mixed properties and become non-propagating as well (for more details see \citealp{2021arXiv210412652Z}). However, these regimes are not a subject of the current study. Thus, to find roots that can be associated with propagating MA waves, we look for wavenumbers in the form $k=k_\mathrm{Re}+\mathrm{i} k_\mathrm{Im}$ and assume that the corresponding frequency is some real non-zero quantity $\mathrm{Re}~\omega \neq 0$. Following this assumption, we can find the exact solution of the dispersion relation (Equation \ref{dispRelation}) as:
\begin{eqnarray}
k_\mathrm{Re}=\sqrt{\frac{\mathrm{Re}\left(r_\pm\right)+\mathrm{Abs}\left(r_\pm\right)}{2}},\quad k_\mathrm{Im}=\frac{\mathrm{Im}\left(r_\pm\right)}{2k_\mathrm{Re}}, \quad r_\pm=\frac{-b\pm\sqrt{b^2+4ac}}{2a}.
\label{solution}
\end{eqnarray}
This solution (Equation~\ref{solution}) describes dispersion properties of slow (root $r_+$) and fast (root $r_-$) waves propagating in the plasma flux tube in the presence of thermal misbalance. Further, we will concentrate on the solution corresponding to the propagating slow waves only, since consideration of fast waves requires additional information about the external medium.

As the final matter of this section, it is worth recalling some applicability condition of  thin-flux-tube approximation  to determine the wave periods of interest.  According to \cite{Zhugzhda96}, the  second-order thin-flux-tube approximation is applicable only when wave period $P$ is greater than the mechanical equilibration time in the radial direction $\left[\tau_\mathrm{R}\right]$. Thus, the subsequent analysis refers to periods $P>\tau_\mathrm{R}$, where $\tau_\mathrm{R} = R/\mathrm{min}\left(c_\mathrm{A},c_\mathrm{S},c_{\mathrm{S}Q}\right)$.

\section{Slow Waves in Active Region Fan Loops }\label{s:WavePrppagation} 

\subsection{Heat-Loss Model}\label{ss:HLmodel}
In what follows, we will consider the slow-wave propagation inside active region fan loops. In order to account for the thermal-misbalance influence on slow waves, one should determine the  heat-loss function  $\left[Q\!\left(\rho, T\right)\right]$. In this study, the following form of the loss part due to the optically thin radiation is used
\begin{equation}
\label{loss_f}
L\!\left(\rho,T\right) = \frac{\rho}{4 m^2}\,\Lambda\!\left(T\right)\,,
\end{equation}
where $m =0.6\times1.67\times10^{-24}\,\mathrm{g}$ is the mean particle mass, $\Lambda\!\left(T\right)$ is the radiative-loss function determined from the CHIANTI atomic database v. 10.0 (\citealp{Dere1997,Delzanna2020chianti}). The heating function $H\!\left(\rho, T\right)$ can be locally modeled as
\begin{equation}
\label{heat_f}
H\!\left(\rho, T\right)=h\rho^aT^b\,,
\end{equation}
where $h$, $a$, and $b$ are given constants. The first constant $h$ is determined from the steady-state condition $Q\!\left(\rho_0,T_0\right)=0$: $h=L\!\left(\rho_0,T_0\right)/\rho_0^a T_0^b$. The power-law indices $a$ and $b$ could be associated with some specific heating mechanism. More frequently, the following five mechanisms are considered (\citealp{Rosner1978,Dahlburg1988,Ibanez1993}): i) constant heating per unit volume ($a=-1$, $b=0$); ii) constant heating per unit mass ($a=0$, $b=0$); iii) heating by coronal current dissipation ($a=0$, $b=1$); iv) heating by Alfv{\'e}n mode/mode conversion ($a=1/6$, $b=7/6$); v) heating by Alfv{\'e}n mode/anomalous conduction dumping ($a=-1/2$, $b=-1/2$). However, it was shown by \cite{Kolotkov_2020} that these mechanisms are incompatible with the observations of widespread coronal thermal stability and the rapid damping of slow (acoustic) waves. In our research, we follow  \cite{Kolotkov_2020} and \cite{ Duckenfield2020effect} and use the values
of $a=1/2$, $b=-7/2$, for which both thermal stability $\tau_{V,P}>0$ and acoustic stability $\left(\tau_{P}-\tau_{V}\right)/\tau_{P}\tau_{V}  >0$ conditions are always satisfied in the coronal plasma. We should mention that the calculations which will be provided further are valid for the chosen heating mechanism. However, the variation of the heating mechanism will lead to the change of the calculated values.

\subsection{Phase Speed of Slow Waves}\label{ss:PhaseSpeed}

In this subsection, we address the period dependence of the slow-mode phase speed in the solar corona. Previously, in Section \ref{s:DispRealation}, we introduced the characteristic timescale $\left[P_\mathrm{m}\right]$ (see Equation~\ref{Per_mis}) indicating the maximum of the dispersion effect caused by the thermal misbalance only. It seems reasonable to introduce the analogous characteristic timescale for the case of geometric-dispersion effect only. The dispersion relation for the case of the thermal-misbalance absence can be obtained  from Equation~\ref{dispRelation} rewritten using the assumption  $ \tau_{P,V} \rightarrow \infty $. Analyzing the obtained dispersion relation, one may estimate that in this case $\mathrm{d} c_\mathrm{ph}/\mathrm{d} \omega$  reaches a maximum near the period $P_\mathrm{g}$:
\begin{equation} \label{Per_geom}
	P_\mathrm{g} = \frac{1}{c_\mathrm{S}+c_\mathrm{T}}\sqrt{\pi A_0 \frac{\left(c_\mathrm{T} +3 c_\mathrm{S}\right)}{\left(c_\mathrm{S}+3 c_\mathrm{T}\right)}  \frac{\left(4 c_\mathrm{A}^2 - \left(c_\mathrm{S}+c_\mathrm{T}\right)^2 \right)}{\left(c_\mathrm{A}^2+ c_\mathrm{S}^2\right)}}, 
\end{equation}
which for low a $\beta$ plasma is reduced to $\approx \pi R / c_\mathrm{S}$.

In the general case, the timescale $P_\mathrm{m}$ may be either greater or smaller than $P_\mathrm{g}$. The relative position of these timescales is completely defined by the loop parameters and the acting heating/cooling mechanisms. As a result, the form of the frequency dependence of the slow-wave phase velocity can vary significantly depending on certain absolute values of timescales $P_\mathrm{m}$ and $P_\mathrm{g}$, as well as their ratio. To demonstrate the  possible thermal misbalance and finite cross-section influence on the slow-wave phase speed for the heat-loss model discussed above (see Section  \ref{ss:HLmodel}), let us consider a fan loop with loop length $L=100\,\mathrm{Mm}$ and radius $R=1\,\mathrm{Mm}$. We also assume that the loop consists of plasma with temperature  $T_0=0.8\,\mathrm{MK}$, number density $n_0=10^{10}\,\mathrm{cm^{-3}}$, and magnetic field $B_0=10\,\mathrm{G}$.

\begin{figure}    
	\centerline{\includegraphics[width=.65\textwidth,clip=]{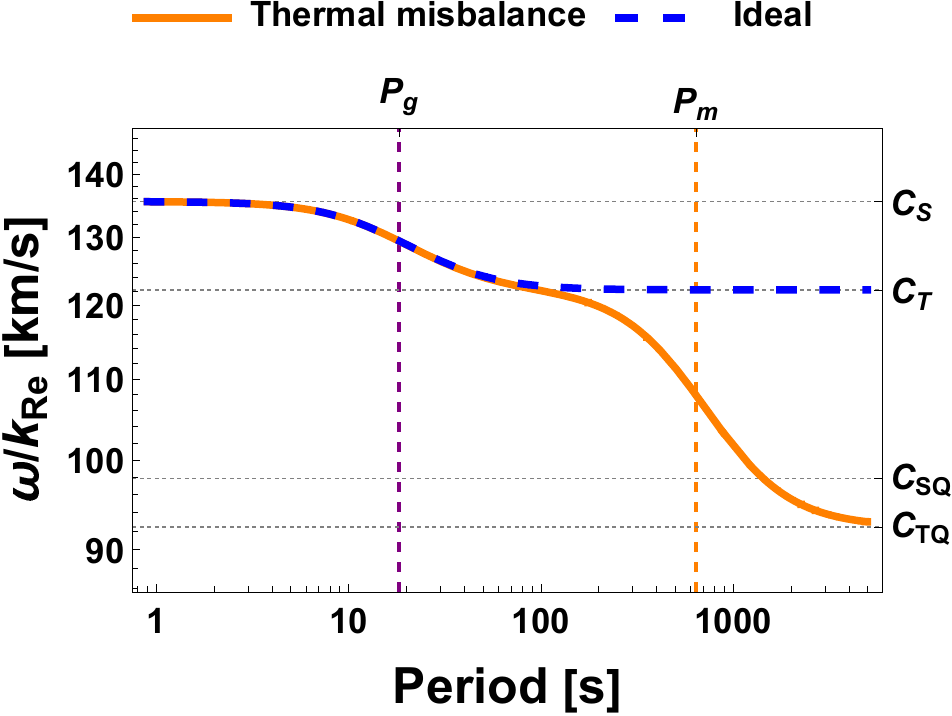}
	}
	\caption{ The phase speeds of slow waves in an active-region coronal loop. The solid-orange line corresponds to the case of combined impact of thermal misbalance and finite cross section. The dashed-blue line corresponds to the case of ideal plasma with the single impact of geometric dispersion. Calculations are made for the fan loop with $L=100\,\mathrm{Mm}$, $R=1\,\mathrm{Mm}$, $T_0=0.8\,\mathrm{MK}$, $n_0=10^{10}\mathrm{cm^{-3}}$, $B_0=10\,\mathrm{G}$, and the discussed heat-loss model (see Section  \ref{ss:HLmodel}). The  mechanical equilibration time in the radial direction $\tau_\mathrm{R} \approx 10$ seconds, for periods less than  $\tau_\mathrm{R}$ the second-order thin-flux-tube approach is not applicable.
	}
	\label{fig-1}
\end{figure}

In Figure \ref{fig-1}, one may find the comparison between phase speeds of slow waves calculated for the case of geometric dispersion only, and the case of combined influence of the thermal misbalance and finite cross section. For the considered heat-loss model (see Section  \ref{ss:HLmodel}), the misbalance timescale is greater than the geometric one: $P_\mathrm{m} > P_\mathrm{g}$. This leads to the fact that for waves with  periods  comparable to or less than $P_\mathrm{g} $, the main impact on dispersion properties comes mostly from the finite cross-section influence. As a result, one can see the change in the phase speed from $c_\mathrm{S}$ to $c_\mathrm{T}$ at these periods, similar to the case of an ideal plasma. Differences in the dependence of the phase speed begin at periods comparable with $P_\mathrm{m}$, since now, in addition to the geometric dispersion, the dispersion due to thermal misbalance plays a role. As a consequence, the value of the phase speed at large periods tends to the value $c_{\mathrm{T}Q}$ (see Equation~\ref{cT0}).

The last feature is quite important for the needs of MHD-seismology, since the long-wavelength limit of the phase speed now is $\lim\limits_{P \rightarrow \infty} c_\mathrm{ph} = c_{\mathrm{T}Q}$,  which may  differ significantly from the generally assumed value  $c_\mathrm{T}$ for an ideal plasma. This issue is discussed in detail in the following section.

\subsection{Deviations from Ideal Plasma Approximation}\label{ss:Errors}

In particular, the above mentioned difference in the phase-speed limit value may be the source of errors in  the seismological estimation of plasma parameters. For example, \cite{Jess2016} measured the loop magnetic field using the slow-wave phase speed $c_\mathrm{ph}$ interpreted as $c_\mathrm{T}$. In this case, the magnetic field is easily determined if the temperature and density of the plasma are known
\begin{equation}
	B_\mathrm{0est}=\sqrt{4\pi\rho_0\frac{c_\mathrm{S}^2 c_\mathrm{ph}^2}{\left(c_\mathrm{S}^2-c_\mathrm{ph}^2\right)}}.
	\label{B0est}
\end{equation}

On the other hand, if the magnetic field and plasma density are known, the temperature can be estimated as follows:

\begin{equation}
T_\mathrm{0est}=\frac{m}{\mathrm{k_B}\gamma}\frac{c_\mathrm{A}^2 c_\mathrm{ph}^2}{\left(c_\mathrm{A}^2-c_\mathrm{ph}^2\right)}.
\label{T0est}
\end{equation} 
\begin{figure}    
	\centerline{\includegraphics[width=.65\textwidth,clip=]{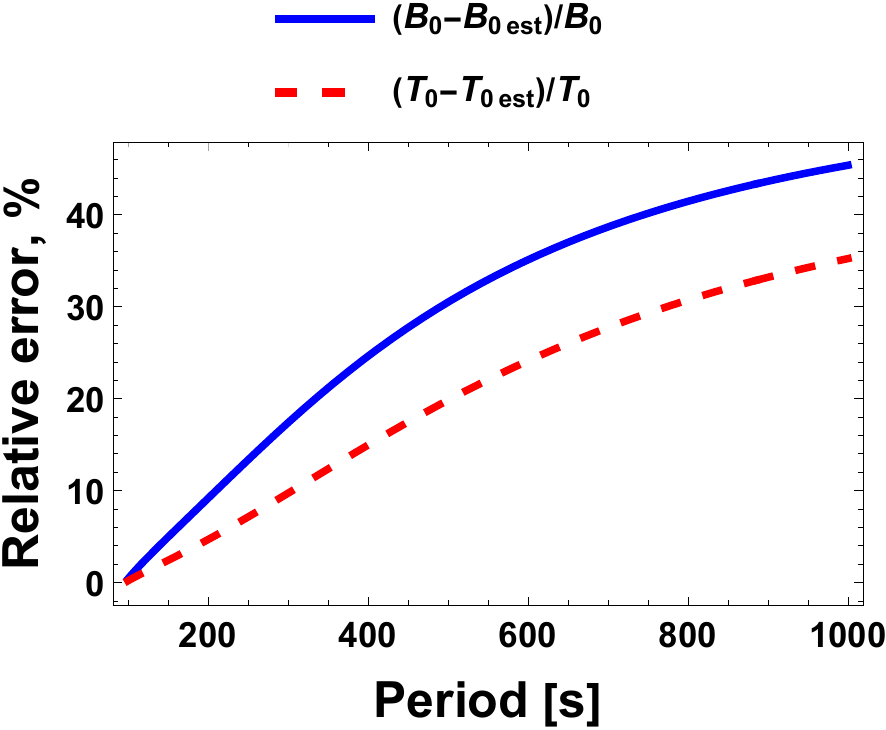}
	}
	\caption{Relative errors in determination of magnetic field strength (solid-blue curve) and temperature (dashed-red curve) caused by neglecting the dispersion of phase velocity associated with thermal misbalance and implying the use of tube speed $c_\mathrm{T}$ as phase speed of slow waves for considered periods.
	}
	\label{fig-2}
\end{figure}
As we show in Figure \ref{fig-1}, the measured phase speed $c_\mathrm{ph}$ for periods $P>P_\mathrm{m}$ may differ significantly from $c_\mathrm{T}$ in the thermally active plasma. Thus, errors in the determination of magnetic field strength and plasma temperature from Equations \ref{B0est} and \ref{T0est} may arise. Figure \ref{fig-2} shows relative errors in determination  of magnetic field strength and temperature by slow waves without accounting for the thermal misbalance, i.e. when measured phase speed $c_\mathrm{ph}$ is interpreted as $c_\mathrm{T}$. It is seen that for a wide range of wave periods, the error can exceed $30\,\%$.

This means that neglecting thermal misbalance may be the reason for the divergence between seismological and spectrometric estimations of plasma parameters. On the contrary, this divergence may be a valuable source for estimations of the coronal heating function $H\!\left(\rho, T\right)$ and its parameters (Equation \ref{heat_f}).

\section{Role of Cross Section and Magnetic Field}\label{s:RoleofAB} 

\subsection{Influence of Cross Section}

	\begin{figure}    
		\centerline{\includegraphics[width=.95\textwidth,clip=]{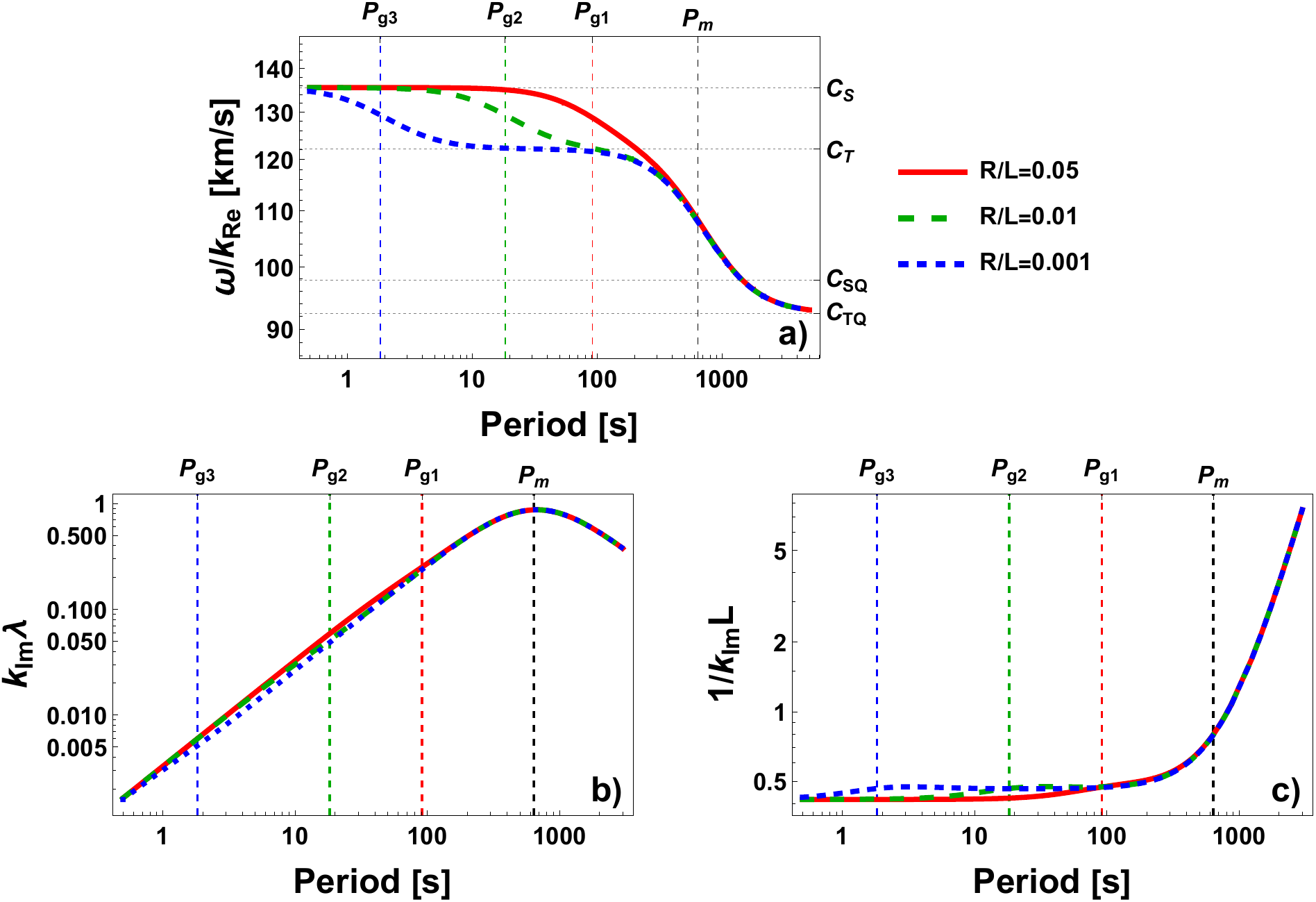}
		}
		\caption{Finite-cross-section influence on the dispersion properties of slow waves. \textbf{a)} Influence on the slow wave phase speed. \textbf{b)}  Influence on the slow-wave quality factor. \textbf{c)} Influence on the slow-wave damping length. Solid-red, medium-dashed-green, and dashed-blue lines as well as characteristic timescales (periods for which geometric dispersion has the most effect) $P_{g1}, P_{g2},$ and $P_{g3}$ correspond to the loop radii $R=0.05\,L,0.01\,L,$ and $0.001\,L$, respectively. The equilibration times in the radial direction $\tau_\mathrm{R}$ calculated for solid-red, medium-dashed-green, and dashed-blue lines $\approx 50, 10,$ and $1$ seconds, respectively
		}	
		\label{fig-3}
	\end{figure}
	
It has been shown in the previous section that the thermal misbalance can noticeably affect the slow-wave phase speed, and it may be crucial for the task of plasma-parameter determination by analysis of observed slow waves. Here, we investigate how the change of tube cross section affects dispersion properties of slow waves in the plasma with thermal misbalance. For this purpose, we use the fan-loop parameters from the previous section, except for the loop radius. We use the following set of the loop radii: $R=0.05\,L,0.01\,L$, and $0.001\,L$. It should be mentioned that various tube radii lead to the different equilibration  times in the radial direction $\left[\tau_\mathrm{R}\right]$ and, as a result, to different ranges of applicability for the thin-flux-tube approximation. 

Figure \ref{fig-3}a demonstrates how the change of cross section affects the phase speed of slow waves. In the range of periods where the impact of thermal misbalance is weak ($P<P_\mathrm{m}$), one can see that the growth of the loop radius leads to the growth of the timescale $P_\mathrm{g}$ associated with the maximum dispersion effect from the geometric dispersion only. However, in the range of periods ($P>P_\mathrm{m}$) with the strong influence of thermal misbalance, the change of loop radius does not lead to any visible effect. Moreover, one can see that for wave periods greater than 300 seconds, which is even less than $P_\mathrm{m}$, the phase speed becomes independent of the cross-section value. 

Figures \ref{fig-3}b and \ref{fig-3}c show the cross section influence on the slow-wave quality factor $\left[k_\mathrm{Im}\lambda\right]$ and on the damping length measured in loop lengths ($\left[1/k_\mathrm{Im}L\right]$). The heat-loss model considered in this article (see Section  \ref{ss:HLmodel}) implies the damping of slow waves. All quality-factor curves have a maxima around $P_\mathrm{m} \approx 640\,$seconds regardless of the cross section. It is clearly seen that for the considered loop parameters the wave damping comes primarily from the thermal misbalance, and the influence of geometry on damping is negligible. 

Estimated damping lengths (Figure \ref{fig-3}c) increase with the wave period. A similar behavior was obtained by \cite{Prasad2014} for slow waves observed in on-disk loop structures. The minimal damping length from Figure \ref{fig-3}c is about $0.44\,L$, which for $L=100\,\mathrm{Mm}$ gives the value of $44\,\mathrm{Mm}$. This value is close to the value of $20\,\mathrm{Mm}$ measured by \cite{Marsh2011} for slow-mode wave propagation in an active-region loop system. However, this value was obtained for $P=12\,$minutes. In the case considered here, the heating mechanism gives the value of $88\,\mathrm{Mm}$. At the same time, the minimum damping length, theoretically estimated by \cite{Marsh2011}, was determined by the loop-area divergence and had the value of $105\,\mathrm{Mm}$ (for parameters from \textit{Solar TErrestrial RElations Observatory}-A (STEREO-A)) or $65\,\mathrm{Mm}$ (for parameters from STEREO-B). It means that the thermal misbalance has the potential to explain the observed value and frequency behavior of slow waves damping lengths. On the other hand, observations of slow-wave damping may provide useful information for estimating the parameters of the coronal heating function. 

\subsection{Influence of Magnetic-Field Strength}

Here, let us investigate the  magnetic-field influence on the dispersion properties of slow waves.  As the considered heating rate (see Equation~\ref{heat_f}) is assumed to be a function of density and temperature only (the case of heating rate additionally proportional to the magnetic field has been considered by \cite{Duckenfield2020effect}), then the magnetic field will affect the dispersion properties only under the terms with the Alfv{\'e}n speed $c_\mathrm{A}$.  For analysis, we will use the loop parameters from Section \ref{ss:PhaseSpeed} with $T_0=0.8\,\mathrm{MK}$, $B_0=5, 10$, and $30\,\mathrm{G}$.

\begin{figure}    
	\centerline{\includegraphics[width=.95\textwidth,clip=]{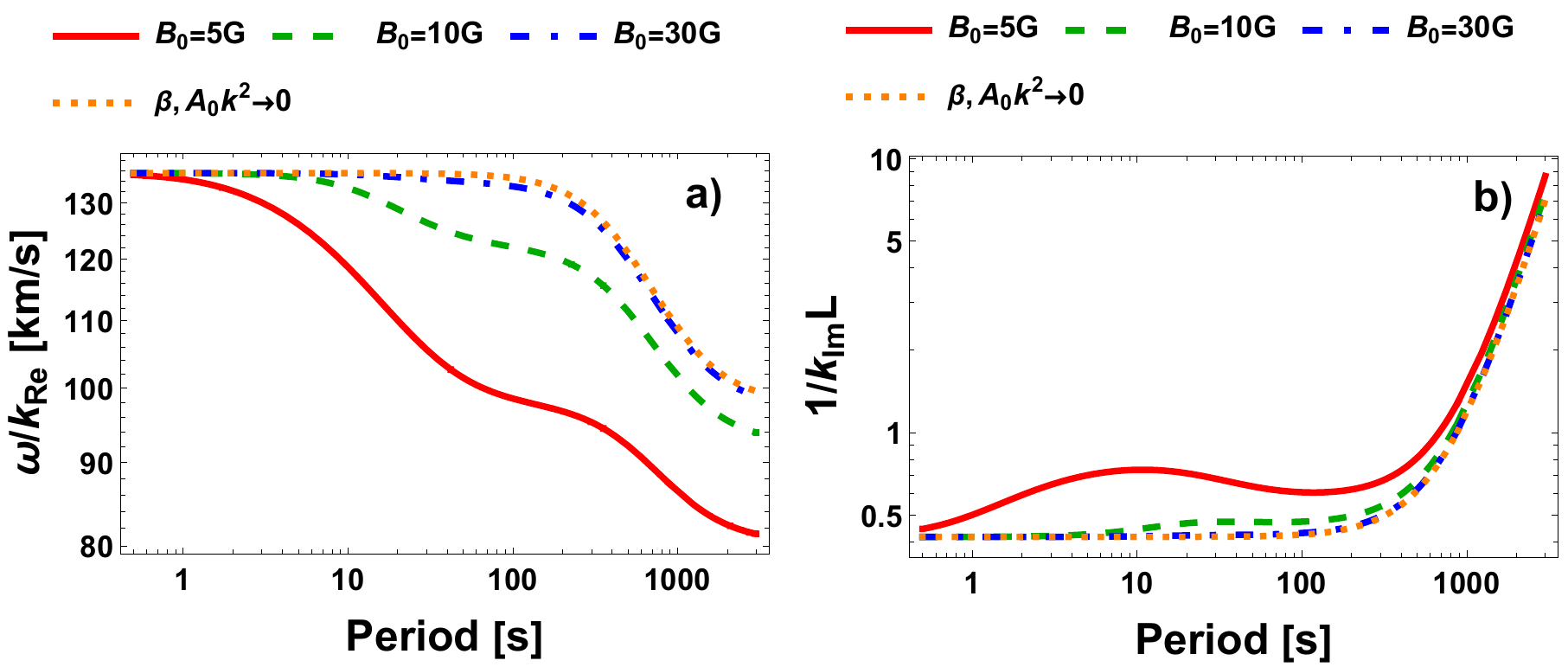}
	}
	\caption{Magnetic-field influence on the dispersion properties of slow waves. \textbf{a)} Influence on the slow-wave phase speed.  \textbf{b)}  Influence on the slow-wave damping length. The solid-red, dashed-green, and dot-dashed-blue lines correspond to $ 5, 10,$ and $30\,\mathrm{G}$, respectively. The  dotted-orange lines indicate dispersion curves obtained applying the infinite field approximation ($\beta \rightarrow 0$ and $A_0 k^2 \rightarrow 0$).
	}
	\label{fig-4}
\end{figure}

Figure \ref{fig-4} shows how the  phase speed and damping length of slow waves vary with the magnetic-field strength. In the strongly non-adiabatic limit (i.e. infinitely long period) the phase speed tends to the modified tube speed $c_{TQ}$. The solid-red and dashed-green curves are far departed from the infinite field limit (orange-dashed line) since the change of magnetic field leads to the change of the Alfv\'en speed $c_\mathrm{A}$ which subsequently causes the change of the modified tube speeds $c_{TQ}$. With  the increase of magnetic field, $c_{TQ}$ tends to $c_{SQ}$. It is seen that for $B_0=5\,\mathrm{G} (\beta \approx 1)$ the impact of the geometric dispersion is significant for the period dependencies of phase speed and damping length and is comparable to the effect of the thermal misbalance. However, with the growth of the magnetic field, the impact of geometric dispersion rapidly decays. One may notice that for $B_0=30\,\mathrm{G} (\beta \approx 0.03)$, the dispersion properties of slow waves are practically indistinguishable from their properties for the case of the uniform plasma and infinite magnetic-field approximation. Thus, for typical coronal loops with the magnetic field $B_0 \gtrsim 30\,\mathrm{G} $, the approximation of infinite magnetic field is quite applicable for the description of slow waves.

\section{Discussion and Conclusion}\label{s:Discussion}

In the current research, the combined influence from thermal misbalance and finite cross section of the loop on the dispersion properties of propagating slow waves has been investigated. The analysis has been conducted with the help of the second-order thin-flux-tube approximation. A linearising the basic system of MHD equations allowed us to obtain evolutionary Equation \ref{waveEq} for compression modes in the plasma. Analysis of this equation reveals the fact that the thermal misbalance widens the range of geometric dispersion's impact. This, in particular, leads to the requirement for the introduction of the modified tube speed $c_{\mathrm{T}Q}$, which is a consequence of both thermal-misbalance and finite-cross-section effects. Investigation of slow-mode dispersion relation and analysis of the period dependence allows us to demonstrate the features mentioned (see Figure \ref{fig-1}).

 Comparison of the phase-speed period dependencies calculated for cases of presence and absence of the thermal misbalance shows that using $c_\mathrm{T}$ as the long-period limit value may be the source of errors in  the seismological estimation of plasma parameters. Looking at Figure \ref{fig-2}, one can see that   for a wide range of wave periods, the error in estimating the magnetic field strength can exceed $30\,\%$ for the chosen heating mechanism.
 
  We also analyze the role of the cross section value and the magnetic-field strength on the dispersion properties of the slow waves for the heating mechanism considered. It is shown, using the heat-loss model presented in Section \ref{ss:HLmodel}, that an increase of the cross section leads to the increase of the period of the maximum geometric dispersion effect. However, it does not affect the periods strongly affected by thermal misbalance $P\gtrsim P_\mathrm{m}$ (see Figure \ref{fig-3}a). Variations of the cross-section do not significantly affect the period dependence of the quality factor and damping length (see Figure \ref{fig-3}b and \ref{fig-3}c, respectively). The impact of the magnetic-field strength is shown in Figure \ref{fig-4}. One may see, that for the magnetic-field strength implying plasma $\beta \approx 1 $, the effect can be considerable. However, in the case of low-$\beta$ plasma (e.g. in our calculations $B_0 > 30\,\mathrm{G}$, i.e. $\beta \leq 0.03$), the effect of geometric dispersion becomes negligible and allows the application of the infinite-field approximation for describing the slow waves.
  
  In conclusion, we want to emphasize that the constructed theory widens our knowledge about the properties and evolution of slow modes in the solar corona. This further understanding allows us the possibility to use slow waves not only as a tool for the estimation of plasma parameters, but also as a tool for the estimation of the non-adiabatic processes (e.g. for phenomenological determination of unknown coronal-heating mechanisms). {This is possible due to the fact that thermal misbalance influence is sensitive to the choice of the heating mechanism.}

%

%

%


%
\begin{acks}
The study was supported in part by the Ministry of Education and Science of Russia by State assignment to educational and research institutions under Project No. FSSS-2020-0014 and No. 0023-2019-0003, and by RFBR, project number 20-32-90018. CHIANTI is a collaborative project involving George Mason University, the University of Michigan (USA), University of Cambridge (UK), and NASA Goddard Space Flight Center (USA).
\end{acks}

{\footnotesize\paragraph*{Disclosure of Potential Conflicts of Interest}
	The authors declare that they have no conflicts of interest. [Edit as appropriate.]
}
\bibliographystyle{spr-mp-sola}
\bibliography{refs}

\end{article} 
\end{document}